# *Ab initio* and Molecular Dynamics Study of The Nanomechanical Properties of DNA Oligomers

Ilya Yanov and Glake Hill[*]

Computational Center for Molecular Structure and Interactions (CCMSI), Department of Chemistry, Jackson State University, MS 39217, United States
* Corresponding author: glakeh@ccmsi.us

___

**Abstract**
Although a vast amount of experimental information is available on the elongation, compression, and folding of proteins in biochemical processes, very little is known about the real structure and molecular dynamics of DNA at an atomic level. Since this area of work is relatively new, this paper reports the results of computer simulation of elongation and compression of B-DNA structures providing new insights into high- energy forms of DNA implicated in these processes.

___

**Introduction**
DNA is known to be very flexible and there have been various experiments conducted to proof it. They conclude that deformations of the double helix occur in biological environments. Double stranded DNA can be extended to twice its normal length before its base pairs break [1]. The study of DNA elasticity will enable researchers to understand the binding of DNA to drugs, genetic regulation, and compact packaging of DNA in living cells.

The technological potential of DNA devices has created the need for new investigations that can be used for predictions of DNA structures and properties. The experimental methods for the determination of protein structures are very time consuming and they provide information for only approximate equilibrium conformations. Nanoscale size of proposed DNA devices terminates the complexity of its electronic and mechanical properties, also making research more difficult.

Mechanisms of DNA stretching are unknown at an atomic level. Computer simulations have been previously used for nucleic acids to predict their most probable structures and for predicting conformational variations [2]. A double stranded B-DNA structure was used to simulate the process of progressive elongation and compression. Results are dependent on the nature of the assumed force field, on the choice of independent structural variables such as Cartesian coordinates, and the route taken to change conformation using molecular dynamics and energy minimization. Therefore, it is not easy to compare the existing results.

The aim of this paper is to perform gradient minimization and molecular dynamics calculations on DNA fragments at finite temperatures. A comparisons of ab initio calculations on small fragments of DNA and molecular dynamics study of the mechanical properties of large scale DNA molecules will also allow to estimate the applicability of molecular mechanics approaches to the overall structural analyses.

**Methods**
From a DNA fragment, experimental xyz coordinate data was obtained and applied using the following formula:
$$X_{i+1}=X_i[1+(N/L)],$$
where $X_i$= initial x-coordinate of the atoms; N= length in angstroms of the DNA fragment; L= initial length in angstroms of the DNA fragment. The total length of the DNA fragment was approximately 40 angstroms and it contained 12 complimentary base pairs. The separation between two base pairs was found to be approximately 3.33 angstroms. The fragment was elongated from +1 to +15 and compressed from -1 to -15 angstroms. When the DNA structure was fully extended, each base pair was approximately 5 angstroms apart. When fully compressed, this separation amounts to approximately 2 angstroms per base pair. Rigid boundary conditions and freezing boundary residues were placed to obtain more uniform elongation. Using AMBER [3] force field methods, the single point energies and geometry optimization energies were calculated for each DNA fragment. In addition, the average total energies were also calculated. These calculations were carried out at 0K, 25K, 50K and 100K temperatures.

The GAUSSIAN94 [10] package was used for the *ab initio* SCF MO calculations at the Hartree-Fock MP2 levels.

**Results**
*Gradient minimization*
It is worthy to note that the stretched DNA double helices reported in previous computer simulation studies [1, 2, 4] such as the "ribbon" and narrow fiber in B-DNA structure from Lebrun & Lavery, the "S- ladder" by Konrad & Bolonick, and the B- DNA transitions in "hyperfamilies" from Kosikov et al. have common characteristics with our extended forms of B- DNA.

Also experimental investigations revealed some characteristics that are in agreement with this study of DNA deformation. Experimental data obtained from Thundat et al. [5] revealed that the double stranded DNA molecules are occasionally found appearing straightened and stretched in atomic force microscope (AFM) images. These stretched structures were prevalent with long lambda phage DNA [2, 5].

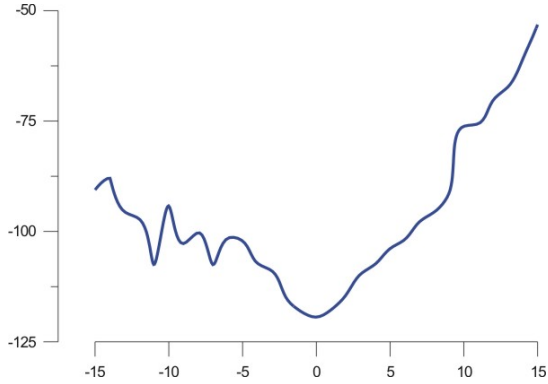

Figure 1. Total energy (kcal/mol) of the DNA fragment vs. elongation (angstrom). AMBER force field model.

Other experimental data performed by Smith et al. [6] showed that when DNA was pulled between two pipettes, maximum extension was 1.85 times the initial length. The maximum geometric extension was approximately 2.1 times its contour length [5], which agrees with Bensimon et al., [7] who reported DNA extension up to 2.1 times of its contour length. Three years later, Cluzel et al. [8] performed DNA extension of approximately 1.7 times of its initial length using the JUMMA [9] computer simulation program. The JUMMA [9] program was used to perform stretching involving energy minimizations as a function of the length of the DNA fragment [8]. These type calculations were conducted in this study using the AMBER [3] force field method.

It is shown (Figure 1) that modulus of elasticity have complex behavior confirming existence of different local minima of total energy. From biochemical point of view, it means that the structure of the DNA may correspond to a metastable state with a long lifetime.

*Molecular dynamics*
For the first time comprehensive molecular dynamics calculations have been performed to determine the temperature dependence of the modulus of elasticity (Figure 2). Our results indicate the significant changes in mechanical properties vs. temperature. It should be noted that the original pattern conserves mostly up to 100K during elongation of DNA, but only up to 25K in the opposite process. Those results also confirm the data obtained by gradient minimization.

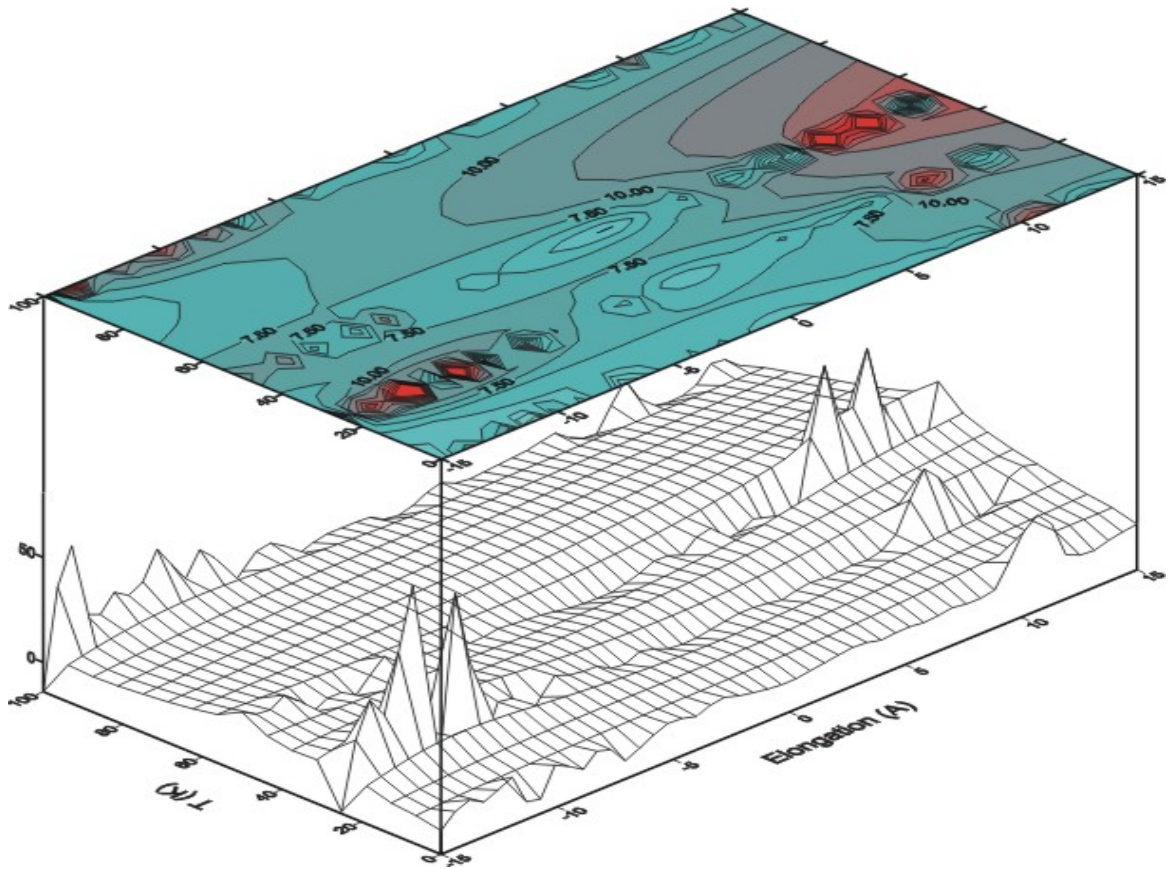

Figure 2. Temperature dependence of the modulus of elasticity.

*Ab initio calculations*

This study allows to explain the nature of the kink (Region II, Figure 1) in the total energy graph. Based on the *ab initio* calculations of DNA base pair stacking interaction one can see asymmetric bell shape energy dependence (Figure 3). The approximate values of stacking energy (10.6 kcal/mol per two base pair or about 60 kcal/mole for 12 base pairs) are in good agreement with the gradient minimization data.

Therefore we can assume that the Region II is related to the base pair stacking interaction.

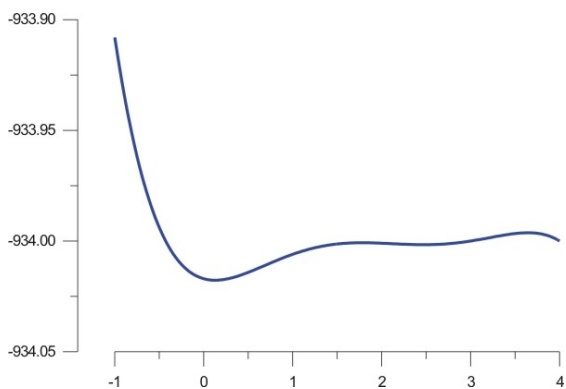

Figure 3. Total energy (a.u.) as the function of the distance (angstrom) between DNA base pairs. MP2/6-31G* level of theory.

**Acknowledgments**

This work was supported by the Department of Defense through the U. S. Army Engineer Research and Development Center (Vicksburg, MS), Contracts #W912HZ-04-2-0002 and #W912HZ-05-C-0051, and NSF-PREM Grant # DMR-0611539.